\documentclass[prl,aps,11pt,preprintnumbers,superscriptaddress]{revtex4}

\usepackage{amsmath,amssymb,amsfonts,amsthm}
\usepackage{amsbsy} 
\usepackage{epsfig}
\usepackage{latexsym}
\input amssym.def
\input amssym.tex

\usepackage{color}
\usepackage{hyperref}

\def\barr{\begin{array}}
\def\earr{\end{array}}

\def\ben{\begin{equation}}
\def\een{\end{equation}}
\def\bs{\begin{subequations}}
\def\es{\end{subequations}}
\def\bena{\begin{eqnarray}}
\def\eena{\end{eqnarray}}

\begin{document}
\title{Group Field Theory and Loop Quantum Gravity}
\author{\bf Daniele Oriti}
\affiliation{Max Planck Institute for Gravitational Physics (Albert Einstein Institute), Am M\"uhlenberg 1, 14476 Golm, Germany, EU; doriti@aei.mpg.de}

\begin{abstract}
We introduce the group field theory formalism for quantum gravity, mainly from the point of view of loop quantum gravity, stressing its promising aspects. We outline the foundations of the formalism, survey recent results and offer a perspective on future developments.
\end{abstract}

\date{\today}

\maketitle

\section{GFT from LQG perspective: the general idea}
In this contribution, we introduce the group field theory (GFT) formalism for quantum gravity \cite{GFT}, mainly from the point of view of loop quantum gravity, arguing why it represents, in our opinion, a most promising setting for future developments. 
We describe the kinematical Hilbert space and its relation to the LQG one, and how the GFT quantum dynamics connects to the canonical one as well as completes spin foam models. We also discuss the problem of defining the continuum limit of such theories and of extracting effective continuum physics, highlighting the important role that GFTs can play in this respect. 
This is not an in-depth introduction, nor a complete review of the literature. We only outline the foundations of the formalism, survey recent results and offer a perspective on future developments.

\

{\bf An historical prelude -}
Group field theories can be approached from different angles, coming from different lines of research in quantum gravity. 
Historically, their first appearance \cite{boulatov, ooguri} came as a development of tensor models \cite{tensorOld} (themselves a generalisation of matrix models \cite{matrix}, which provided a successful quantisation of (pure) 2d gravity), allowing to make contact with state sum formulations of 3d quantum gravity (Ponzano-Regge and Turaev-Viro model), whose relation with simplicial quantum gravity, e.g. quantum Regge calculus \cite{RuthRegge}, was already known, and more generally topological BF theory in any dimension. These first models were obtained by taking the simplest tensor model for 3d simplicial gravity and: 1) replacing the domain set for the tensor indices with a group manifold ($SU(2)$); 2) adding a gauge invariance property to the field (tensor), with the effect of introducing a gauge connection on the lattices generated by the perturbative expansion of the model. The triviality of the kinetic and interaction kernels (simple delta functions on the group) in the GFT action resulted in the amplitudes being exactly those of BF theory discretised on the same lattices (imposing flatness of the connection). Written in terms of group representations, the same amplitudes took the form of the mentioned state sums. This is the first way to understand group field theories: GFTs can be seen as tensor models enriched by algebraic data with a quantum geometric interpretation (allowing a nice encoding of discrete gravity degrees of freedom), or, equivalently, as more general class of combinatorially non-local field theories of tensorial type. The relation between state sum models of topological field theory, and their GFT formulation, and loop quantum gravity was soon pointed out in \cite{CarloBoulatov} (where the link to the dynamical triangulations approach \cite{DT} was also mentioned): the boundary states of such models matched the newly developed loop representation for quantum gravity \cite{LQG}. Indeed, spin networks (introduced in LQG immediately afterwards) describe also the Hilbert space of GFT models. The latter acquire then a nice interpretation from the LQG perspective: GFTs are quantum field theories for spin networks, providing them with a covariant dynamics. This covariant definition started being developed a few years later \cite{DPFKR,P-R,mikecarlo}. Indeed, more interest in GFTs as quantum gravity models came with the realisation that they provide a complete definition of spin foam models for 4d gravity \cite{SF}: they capture the same quantum amplitudes as Feynman amplitudes, organising them coherently in a sum over spin foam complexes, arising from the perturbative expansion of the field theory. It is in this context that most developments have taken place in the following years, bringing in particular an improved understanding of the quantum geometry encoded in several interesting spin foam models for 4d gravity. Finally, in recent years we have witnessed a renaissance of tensor models \cite{tensor}, with important mathematical results concerning their combinatorial structures. Since they share the same combinatorial structures, this has triggered further developments in GFTs. Indeed, it is by combining the tools coming from tensor models and the quantum geometric understanding provided by loop quantum gravity and spin foam models that, we believe, we have the best chance to make progress on the remaining open issues of quantum gravity, within the GFT framework.

\

\noindent {\bf A brief definition -} A (single-field) GFT is a theory of a field $\varphi : G^{\times d} \rightarrow \mathbb{C}$ defined on $d$ copies of a group manifold $G$, with action 
\begin{eqnarray}
S(\varphi,\varphi^*)\,&=&\, \int [dg_I] [dg'_J] \varphi^*(g_I) \mathcal{K}(g_I,g'_J)\,\varphi(g'_J) \, +\, \nonumber \\ &+&\,\sum_i \frac{\lambda_i}{D_i!} \int [dg_{I1}]...[dg_{JD_i}] \varphi^*(g_{I1})\dots\mathcal{V}_i(g_{I1},...,g_{JD_i}) \cdots \varphi(g_{JD_i}) 
\end{eqnarray}
where $i$ labels the possible interaction terms (weighted by coupling constants), each involving $D_i$ fields (or their complex conjugates), in turn depending on $d$ group elements each. A specific GFT model is defined by a choice of group $G$, dimension $d$, kinetic  and interaction kernels $\mathcal{K}$ and $\mathcal{V}_i$. The crucial feature of GFT models, as opposed to ordinary (local) QFTs on space-time, beside the physical interpretation of all the ingredients, is that, the interaction kernels, field arguments are related in a {\it combinatorially non-local} way, i.e. each field is correlated to the others only through {\it some} of its arguments. The specific combinatorial pattern of such correlations is another defining property of specific models, as we will discuss (see \cite{GFTallLQG} for an extensive treatment of the combinatorial aspects of GFTs and their extension to combinatorial structures of arbitrary valence).

This combinatorial non-locality becomes manifest in the quantum theory, defined (in perturbation theory) by the partition function:
\begin{equation}
Z\,=\,\int\mathcal{D}\varphi\mathcal{D}\varphi^*\,e^{-\,S(\varphi,\varphi^*)}\,=\, \sum_\Gamma\frac{ \prod_i\lambda_i^{n_{i}(\Gamma)}}{sym(\Gamma)}\,\mathcal{A}_\Gamma \label{GFTPartFunc}
\end{equation}
where $\Gamma$ denotes GFT Feynman diagrams, $sym(\Gamma)$ the order of their automorphism group, $n_i(\Gamma)$ the number of interaction vertices of type $i$, and $\mathcal{A}_\Gamma$ is the corresponding Feynman amplitude, obtained as usual by convolution of interaction kernels with propagators (inverse of kinetic terms). Again, the explicit form of the Feynman amplitudes and the exact combinatorial structure of the diagrams $\Gamma$ depend on the specific model considered. The combinatorial non-locality of the interaction terms, however, generically implies that GFT Feynman diagrams are not graphs but cellular complexes of arbitrary topology.

The combinatorial structure of GFT fields, action, Feynman diagrams, quantum states and amplitudes, are the same as those of tensor models \cite{tensorOld,tensor}
. However, GFTs {\it enrich} tensor models by the additional group-theoretic data. Because of them, GFTs are quantum field theories whose basic quanta are spin network vertices, i.e. representable as nodes with d open links, labelled by the same algebraic data of LQG states, and their Feynman amplitudes $\mathcal{A}_\Gamma$ for any interaction process $\Gamma$ of these quanta are generically spin foam amplitudes. 
Spin foam amplitudes are, in turn, dual to lattice gravity path integrals, so that GFTs combine also the main idea of dynamical triangulations (quantum gravity as a sum over random lattices) \cite{DT} and the main idea of quantum Regge calculus\cite{RuthRegge} (quantum gravity as a sum over geometric data assigned to a give lattice).

In the following we will highlight structures and concepts shared with other ways of doing loop quantum gravity, as well as points of departure and new concepts brought in by the GFT reformulation.
We will also discuss how GFTs cast the problem of defining a background independent theory of quantum gravity based on LQG ideas in a more or less standard QFT language. This allows the use of several powerful tools, to realise concretely the suggestive notion of \lq atoms of quantum space\rq and to treat spacetime, indeed, like a condensed matter (or many-atom) quantum system, suggesting new lines of developments.

\section{GFT kinematics: Hilbert space and observables}
\noindent {\bf Fock space of quantum states -}
The Hilbert space of states for single-field GFTs is a Fock space built out of a fundamental \lq single-atom\rq ~Hilbert space $\mathcal{H}_v = L^2(G^{\times d})$:
$\mathcal{F}(\mathcal{H}_v) = \bigoplus_{V=0}^\infty\,sym\left\{\left( \mathcal{H}_v^{(1)} \otimes \mathcal{H}_v^{(2)} \otimes \cdots \otimes \mathcal{H}_v^{(V)}\right)\right\}$, where $sym$ indicates symmetrisation with respect to the permutation group $S_V$\cite{GFT-LQG}. This encodes a bosonic statistics for field operators (other possibilities can be considered \cite{GFTdiffeo,colouredGFT}, but they have not been used in the spin foam and LQG context):
\begin{equation}
\left[ \hat{\varphi}(\vec{g})\,,\,\hat{\varphi}^\dagger(\vec{g}') \right]\,=\, \mathbb{I}_G(\vec{g}, \vec{g}') \hspace{0.5cm} \left[ \hat{\varphi}(\vec{g})\,,\,\hat{\varphi}(\vec{g}') \right] = \left[ \hat{\varphi}^\dagger(\vec{g})\,,\,\hat{\varphi}^\dagger(\vec{g}') \right] \,=\, 0
\end{equation}
where $\mathbb{I}_G(\vec{g}, \vec{g}')\equiv  \prod_{i=1}^d\delta(g_i (g_i')^{-1})$, and we used the notation $\vec{g} = (g_1,..,g_d)$. 

In quantum gravity models the group $G$ is chosen to be the local gauge group of gravity in the appropriate space-time dimension and signature, i.e. $G= SU(2),SL(2,\mathbb{R})$ in 3 dimensions and $G= Spin(4),SL(2,\mathbb{C})$ in dimension 4 (or their rotation subgroup $SU(2)$, in order to connect with LQG). 

Each Hilbert space $\mathcal{H}_v$ provides the space of states of a single "quantum" of the GFT field, a quantum gravity \lq atom\rq. It can be understood as a fundamental spin network vertex, represented by a node with $d$ outgoing links (ending up in 1-valent nodes), labelled by group elements, or as a 3-cell (polyhedron) with $d$ boundary faces. This just a pictorial representation. Whether the states represent quantum gravity spin network vertices or geometric polyhedra depends on the type of data they carry and the dynamics they satisfy. For $G=SU(2)$, and with the closure condition $\varphi(g_I) = \varphi(h g_I) \qquad \forall h \in G$ imposed on the fields, however, the polyhedral interpretation is justified and the same is true for $G=SL(2,\mathbb{C})$ and $G=Spin(4)$ with simplicity constraints and closure conditions correctly imposed. In particular, for $d=4$, the GFT quanta represent quantum tetrahedra, about which a lot is known in the spin foam literature \cite{tetrahedron}. In this last case, the basic Hilbert space is $\mathcal{H}_v = \bigoplus_{J_i\in \mathbb{N}/2} Inv \left( \mathcal{H}^{J_1} \otimes ...\otimes \mathcal{H}^{J_4}\right)$, where each $\mathcal{H}^{J_i}$ is the Hilbert space of an irreducible unitary representation of $SU(2)$ labeled by the half-integer $J_i$.    

\

\noindent {\bf Quantum observables -} Kinematical observables are functionals of the field operators $\mathcal{O}\left( \hat{\varphi},\hat{\varphi}^\dagger\right)$. Of special importance are polynomial observables, whose evaluation in the vacuum state defines to GFT n-point functions\cite{alexcarlo}. Any convolution of a finite number of GFT field operators with appropriate kernels would define one such observable, as in any quantum field theory. The peculiarity of GFTs, with respect to ordinary QFTs, is the possibility for these kernels to have a richer combinatorial structure, involving a non-local pairing of field arguments, i.e. relating only a subset of the $d$ arguments of a given GFT field  with a subset of the arguments of a different one. 
Of particular interest for LQG are \lq spin network observables\rq: \begin{equation}O_{\Psi=(\gamma,
  J_{(ij)}^{(ab)},\iota_{i})}(\hat{\varphi}^\dagger)=\left(\prod_{(i)}\int [dg_{ia}]\right)
\Psi_{(\gamma, J_{(ij)}^{(ab)},\iota_i)}(g_{ia}g_{jb}^{-1})\prod_i \hat{\varphi}^\dagger(g_{ia}), \label{spinnet}\end{equation}
where  $\Psi_{(\gamma, J_{(ij)}^{(ab)},\iota_i)}(G^{ab}_{ij})$ identifies a spin network
functional labelled by
a closed graph $\gamma$ with representations $J_{(ij)}^{(ab)}$ associated to the different edges linking two vertices $i$ and $j$,
and intertwiners $\iota_i$ associated to its vertices; $g_{ia}$ (resp. $g_{jb}$) (with $a,b=1,...,d$) are
group elements being the arguments of the field associated to the vertex $i$ (resp. $j$), so that a pair of indices $(a,b)$ denotes each of the edges connecting two vertices $i$ and $j$. The bosonic statistics implies a symmetrisation of $\Psi$ with respect to permutations of the vertex labels. These observables act on the Fock vacuum creating a spin network state associated to a graph $\gamma$. 

\

\noindent {\bf GFT as 2nd quantised reformulation of the LQG kinematics -} We now discuss in more detail in what sense GFT provides a 2nd quantised formalism for spin networks and how one can link (a certain version of) canonical LQG and GFT directly, without passing through the spin foam formulation, but providing in turn a clear link between the latter and canonical LQG. More details can be found in \cite{GFT-LQG} .

By \lq LQG kinematical Hilbert space\rq ~we intend, here, a Hilbert space constructed out of states associated to closed graphs and such that, for each graph $\gamma$, we have $\mathcal{H}_\gamma = L^2\left( G^E/G^V, d\mu =\prod_{e=1}^E d\mu_e^{Haar}\right)$ (here $G=SU(2)$), where $e$ are the links of the graph ($E$ is their total number), with a graph-based scalar product defined the Haar measure on each link $\mu_e^{Haar}$. 
The same Hilbert space can be represented also in the flux basis, via the non-commutative Fourier transform \cite{flux,QuantFlux}, in terms of functions of Lie algebra elements, that are the natural \lq momentum\rq ~variables for the classical LQG phase space on a given graph: $\left[\mathcal{T}^*G\right]^{\times E}$ (before constraints).
The union for all graphs of such Hilbert spaces is, of course, not a Hilbert space. In the LQG and spin foam literature, one finds different ways in which these graph-based Hilbert spaces can be organised to define the Hilbert space of the theory. One is to simply consider the direct sum over all possible graphs: $\mathcal{H}_{LQG}^1 = \oplus_{\gamma}\mathcal{H}_\gamma$. Another, corresponding to the canonical construction in the continuum, is to define appropriate equivalence classes for states over different graphs and then take the projective limit of infinitely refined graphs: $\mathcal{H}_{LQG}^2 = \text{lim}_{\gamma\rightarrow\infty}\frac{\cup_\gamma \mathcal{H}_\gamma}{\approx}$. Of course, the two spaces are very different. The GFT Hilbert space can be understood as a different proposal to define a Hilbert space out of a union of the graph-based Hilbert spaces, by \lq decomposing them into elementary building blocks\rq .    

The basic idea is to consider any wave function in $\mathcal{H}_\gamma$, where $\gamma$ is a graph with $V$ nodes, as an element of $\mathcal{H}_V = L^2\left( (G^{\times d}/G)^{\times V} , d\mu = \prod_{v=1}^V\prod_{i=1}^d d\mu_{Haar,i}^v\right)$, satisfying special restrictions. The latter space can be understood as the space of $V$ spin network vertices, each possessing $d$ outgoing open links, and the extra restrictions enforce the gluing of suitable pairs of such open links to form the links of the graph $\gamma$. In group space, these extra restrictions are conditions of invariance under the group action, which can be enforced through projectors. A function $\Psi_\gamma$ can be obtained from a wavefunction $\phi_V\in\mathcal{H}_V$ as
\begin{equation}
\Psi_\Gamma(G_{ij}^{ab})\,=\,\prod_{[(ia),(jb)]}\int_{G}d\alpha_{ij}^{ab}\,\phi_V(\ldots,g_{ia}\,\alpha_{ij}^{ab},\ldots,g_{jb}\alpha_{ij}^{ab},\ldots)\,=\,\Psi_\Gamma(g_{ia}
(g_{jb})^{-1})\,, \label{gluingGroup} \end{equation}
with the same notation as in ~\ref{spinnet}. 
This defines an embedding of elements of $\mathcal{H}_\gamma$ into $\mathcal{H}_V$. The same construction can be phrased in the flux and spin representations. Moreover, the scalar product of two quantum states in $\mathcal{H}_V$ associated to the same graph agrees with the one computed in $\mathcal{H}_\gamma$ (i.e. the scalar product in $\mathcal{H}_V$, once restricted by gluing conditions associated to the graph $\gamma$, reduces to the one in $\mathcal{H}_\gamma$). This means that $\mathcal{H}_\gamma$ is embedded faithfully in $\mathcal{H}_V$.  Obviously $\mathcal{H}_V$ also contains states associated to open graphs, that is graphs with some links ending up in 1-valent vertices, i.e. with links of open spin network vertices not glued to any other.

The physical picture behind $\mathcal{H}_V$ is that of a \lq many-atom\rq ~Hilbert space, with each \lq quantum gravity atom\rq ~corresponding to a Hilbert space $\mathcal{H}_v = L^2\left( G^{\times d}/G\right)$.  An orthonormal basis $\psi_{\vec{\chi}}(\vec{g})$ in each $\mathcal{H}_v$ is given by the spin network wave functions for individual spin network vertices (labelled by spins and angular momentum projections associated to their $d$ open edges, and intertwiner quantum numbers):
\begin{equation}
\vec{\chi} = \left( \vec{J}, \vec{m}, \mathcal{I} \right) \;\;\;\rightarrow\;\;\; \psi_{\vec{\chi}}(\vec{g}) =  \langle \vec{g}|\vec{\chi}\rangle = \left[ \prod_{a=1}^{d} D^{J_a}_{m_a n_a}(g_{a})\right]\,C^{J_1...J_d,\mathcal{I}}_{n_1..n_d} \;\;\; .
\end{equation}

The Hilbert space is then extended to include arbitrary numbers of QG atoms $\mathcal{H}_{GFT} = \bigoplus_{V=0}^\infty \mathcal{H}_V$
and can be turned into a Fock space by standard methods \cite{GFT-LQG}
introducing the fundamental GFT field operators 
\begin{equation}
\hat{\varphi}(g_1,..,g_d)\equiv\hat{\varphi}(\vec{g})=\sum_{\vec{\chi}}\;\hat{\varphi}_{\vec{\chi}}\; \psi_{\vec{\chi}}(\vec{g}) \hspace{1cm}
\hat{\varphi}^\dagger(g_1,..,g_d)\equiv\hat{\varphi}^\dagger(\vec{g})=\sum_{\vec{\chi}}\; \hat{\varphi}^\dagger_{\vec{\chi}} \;\psi^*_{\vec{\chi}}(\vec{g}) \;\;\; \nonumber ,
\end{equation}
satisfying the commutation relations introduced above. The choice of bosonic statistics, we stress again, is, at this stage, an assumption to be better justified. Acting on the Fock vacuum, these operators generate the GFT Fock space already introduced
. 

Similarly, quantum observables can be turned from 1st quantised operators (i.e. operators acting on the many-atom Hilbert spaces $\mathcal{H}_V$) to 2nd quantised operators on the Fock space, following again standard procedures. Given the matrix elements $\mathcal{O}_{n,m}\left( \vec{\chi}_1,...,\vec{\chi}_m, \vec{\chi}'_1,...,\vec{\chi}'_n\right)$ (or the correspondent functions in the group or flux basis) of the relevant operator $\widehat{\mathcal{O}_{n,m}}$ in a basis of open spin network vertices, take the appropriate convolutions of such functions with creation and annihilation operators, according to which spin network vertices are acted upon by the operator and which spin network vertices result from the same action, to obtain its 2nd quantized counterpart. The result will thus be a linear combination of polynomials of creation and annihilation operators, i.e. of GFT field operators, thus a GFT observable:
\begin{eqnarray}
&\widehat{\mathcal{O}_{n,m}}&\;\;\; \rightarrow \;\;\;\langle \vec{\chi}_1, ....,\vec{\chi}_m  | \widehat{\mathcal{O}_{n,m}} | \vec{\chi}'_1 , ... , \vec{\chi}'_n \rangle = \mathcal{O}_{n,m}\left( \vec{\chi}_1,...,\vec{\chi}_m, \vec{\chi}'_1,...,\vec{\chi}'_n\right) \hspace{1cm}  \rightarrow \nonumber \\ &\rightarrow& \widehat{\mathcal{O}_{n,m}}\left(\hat{\varphi},\hat{\varphi}^\dagger\right) = \int  [d\vec{g}_i][d\vec{g}'_j]\; \widehat{\varphi}^\dagger(\vec{g}_1)..\widehat{\varphi}^\dagger(\vec{g}_m)\mathcal{O}_{n,m}\left( \vec{g}_1,..,\vec{g}_m, \vec{g}'_1,..,\vec{g}'_n\right) \widehat{\varphi}(\vec{g}'_1)..\widehat{\varphi}(\vec{g}'_n)\;\;\;\;\nonumber .
\end{eqnarray} 

\

\noindent {\bf Similarities and differences with the LQG Hilbert space -}
The kinematical Hilbert space of GFT is analogous to the one in LQG in the sense that its quantum states are the same type of functions on group manifolds, associated to graphs, and characterised by the same representation labels, group or Lie algebra elements. Thus they also encode quantum gravity degrees of freedom in purely combinatorial and algebraic structures, and we have seen that, when restricting attention to states associated to the same graph, the corresponding Hilbert spaces actually coincide. 
However, there are also key differences. First of all, there is a priori no embedding of GFT states into a continuous manifold of given topology. 
Quantum states of the type we considered, thus, can be associated to abstract graphs, in the spirit of \lq Algebraic LQG\rq \cite{AlgLQG}. This means that there is a priori no action of diffeomorphisms, nor any knotting degrees of freedom. Thus they also differ from the s-knot states of the diffeo-invariant Hilbert space of canonical LQG. The only symmetry follows from choice of quantum statistics, i.e. symmetry under permutations of vertex labellings. From this point of view, the GFT state space takes the combinatorial and algebraic nature of the degrees of freedom of quantum space to be fundamental, and no continuum intuition is assumed. In fact, there is no attempt to define a continuum limit at this kinematical level, if not in the sense of a limit of infinite number of QG atoms (akin to a thermodynamic limit in condensed matter). In particular, no cylindrical equivalence among GFT states is imposed, and graph links labeled with trivial connection or zero representation label are not neglected (as atoms with zero momentum in condensed matter).
Moreover, while GFT states associated to graphs with different numbers of nodes are by definition orthogonal, GFT states associated to different graphs with the same number of nodes are not orthogonal, contrary to LQG states. 
One could say that graph structures are given less relevance than in standard LQG, because they are reduced to specific correlations among the fundamental GFT quanta and quantum states for the same number of quanta but with different correlations can overlap. At the same time, the physical relevance of graph structures is somewhat enhanced, because no link with spin zero is removed and because the number of graph nodes is turned into a new (very simple) physical observable.
Thus we have many similarities (in particular GFT and LQG use the same algebraic data to describe quantum geometries), but also some differences. The motivation to accept these differences and drop some features of the LQG Hilbert space (e.g. those coming from a continuum space-time background) and take seriously the GFT one with its fundamental discreteness is that the GFT Hilbert space has a clear Fock structure, giving straightforward meaning to the notion of \lq QG atom of quantum space\rq, and making powerful analytical tools available.


\section{The quantum dynamics}
Given the general set-up, one needs to specify a GFT model, that is: a group manifold, the valence $d$ of the fields, and a set of kinetic and interaction terms, including their encoded combinatorial patterns.
In quantum field theory, having fixed a topological space-time manifold, the ingredients selecting a specific model are chosen from few basic considerations: a locality principle, symmetries (which determine the field content and the general theory space), simplicity. The renormalization group is then used as a check of consistency, and as a way to generate effective dynamics at different scales. Locality is a fundamental assumption (alongside other axioms), well grounded in phenomenology. Symmetry principles and field content are also usually suggested by phenomenology. But in the GFT case, at present, the only type of \lq indirect phenomenology\rq we can rely on is the current theories of spacetime and fundamental interactions, and the basic insights of other existing attempts at constructing a quantum theory of spacetime.

Indeed, one can identify three main strategies, in the GFT literature, stemming from three main ways to approach the formalism, and three different directions in quantum gravity research, all converging somehow to the GFT formalism. 

\

\noindent {\bf GFT dynamics from canonical LQG -} The first is suggested by the picture of GFTs as 2nd quantised theories of spin networks and of canonical LQG \cite{GFT-LQG}. We have already seen that GFTs take seriously the basic insights of traditional LQG, based on canonical quantisation of GR in the continuum, in particular the same type of quantum states (even if not same Hilbert space). Should we choose also the same dynamics, encoded in some Hamiltonian constraint operator?  One may say that it is not so reasonable to expect that the fundamental dynamics of space-time at the Planck scale (and maybe beyond) is obtained simply by the operator version of the GR dynamics. GR may be, after all, just an effective theory, and indeed it is only tested at scales many orders of magnitude away from the Planckian. In fact, we have already pointed out how in GFTs many ingredients of the LQG state space, inspired directly from the continuum setting, are dropped. However, to derive a canonical quantum dynamics from the continuum classical dynamics is still a valid possibility, and conventional wisdom about effective theories and running of scales may not apply in a background independent context. On the other hand, we have seen that even continuum canonical LQG ends up with discrete, combinatorial, algebraic structures.  One may then take these discrete structures as fundamental, and look for simplest definition of dynamics for them, not necessarily taken from any continuum spacetime dynamics. Whatever route one decides to follow, if one has such a canonical operator dynamics, the GFT reformulation of the same (defining a specific GFT model) is straightforward, at least at a heuristic level.  
 
First of all, one has to find the 2nd quantised counterpart of the canonical dynamical operator. This could be directly an Hamiltonian constraint or a \lq projection\rq operator $\widehat{P}$ onto solutions of the Hamiltonian constraint equation, such that: $\widehat{P} \, | \Psi \rangle \, = \, | \Psi \rangle$. When written as an operator on the Fock space, such operator will decompose into operators whose action involves 2,3,...,$(n+m)$ spin network vertices, weighted by coupling constants. This decomposition may well involve an infinite number of components, and beside symmetry conditions, the physical question is which terms in this decomposition are actually relevant and whether higher order terms can be reduced to lower order ones. 
For each $(n,m)$-body component of $\widehat{P}$, we consider the matrix elements in a complete basis of products of single-vertex states
and construct the 2nd quantised projector operator (choosing normal ordering), acting on the Fock space, constructed from GFT field operators as:
$$
\widehat{F} | \Psi \rangle \equiv \sum_{n,m }^\infty \lambda_{n,m} \;\left[ \sum_{\{ \vec{\chi},\vec{\chi}'\} }\hat{\varphi}_{\vec{\chi}_1}^\dagger ... \hat{\varphi}_{\vec{\chi}_m}^\dagger \; P_{n,m}\left( \vec{\chi}_1,...,\vec{\chi}_m, \vec{\chi}'_1,...,\vec{\chi}'_n\right) \; \hat{\varphi}_{\vec{\chi}_1'} ... \hat{\varphi}_{\vec{\chi}_n'}\; -\;\sum_{\vec{\chi} }\hat{\varphi}_{\vec{\chi}}^\dagger \hat{\varphi}_{\vec{\chi}}\,\right] | \Psi \rangle = \; 0  \;\;\;\;
$$ 
Even given the above GFT operator, the identification of the corresponding GFT action and partition function has to proceed in a rather heuristic manner. One would like to define a partition function $Z$ for the canonical quantum LQG theory, that is for arbitrary states in the Fock space, thus arbitrary collections of spin network vertices (including those associated to closed graphs). The simplest choice would be an analogue of the {\it microcanonical ensemble}, in which only states solving the canonical dynamical equation contribute: 
$Z_m \, =\,  \sum_{s} \langle s | \;\delta( \widehat{F})  | s\rangle$, where $s$ denotes an arbitrary complete basis of states in the Hilbert (Fock) space of the quantum theory. 
The GFT dynamics (of existing GFT models), however, corresponds to a quantum LQG dynamics of a more general type, which amounts to a choice of a density operator of the {\it grandcanonical} type
$$
Z_g \, = \,  \sum_{s} \langle s | e^{-\, \left(\widehat{F} \, -\, \mu \widehat{N}\right)}  | s\rangle \qquad ,
$$
where the sign of the chemical potential $\mu$ determines whether states with many or few spin network vertices are favoured. 
To rewrite the above partition function as a GFT path integral, we introduce a basis of eigenstates of the GFT field operator:
$$
Z_g = \sum_{s} \langle s | e^{-\, \left(\widehat{F} \, -\, \mu \widehat{N}\right)}  | s\rangle \, = \,   \int \mathcal{D}\varphi \mathcal{D}\overline{\varphi}\, e^{-\, |\varphi|^2}\, \langle \varphi |\, e^{-\, \left(\widehat{F} \, -\, \mu \widehat{N}\right)} \,  | \varphi \rangle \qquad .
$$
This is a GFT path integral with quantum amplitude
$
e^{-\, |\varphi|^2}\, \langle \varphi |\, e^{-\, \left(\widehat{F} \, -\, \mu \widehat{N}\right)} \,  | \varphi \rangle \, \equiv\, e^{- \, S_{eff}}
$
where the effective action $S_{eff}$ is obtained from a {\it classical action} $S_0$ as:
$$
S_{eff}\left(\varphi,\overline{\varphi}\right) \, =\, S\left(\varphi,\overline{\varphi}\right)\, +\, \mathcal{O}(\hbar)\, =\, \frac{\langle \varphi | \widehat{F} | \varphi \rangle}{\langle \varphi | \varphi \rangle} \, + \, \mathcal{O}(\hbar) \qquad .
$$ 
Quantum corrections may amount to new interaction kernels or to a redefinition of the coupling constants for the ones in $S$.
For a given operator equation, then, the corresponding classical (and bare) GFT action is of the form:
\begin{eqnarray}
&S\left( \varphi, \varphi^\dagger\right)&\; =\; m^2 \int d\vec{g}\; \varphi^\dagger(\vec{g})\,\varphi(\vec{g})\; - \hspace{12cm} \nonumber \\ &-&\;\sum_{n,m} \lambda_{n+m} \;\left[\int  [d\vec{g}_i]\,[d\vec{g}'_j]\; \varphi^\dagger(\vec{g}_1)...\varphi^\dagger(\vec{g}_m)\;V_{n+m}\left( \vec{g}_1,...,\vec{g}_m, \vec{g}'_1,...,\vec{g}'_n\right) \varphi(\vec{g}'_1)...\varphi(\vec{g}'_n)\right] \quad \\ &{}&V_{n+m}\left( \vec{g}_1,...,\vec{g}_m, \vec{g}'_1,...,\vec{g}'_n\right) = P_{n+m}\left( \vec{g}_1,...,\vec{g}_m, \vec{g}'_1,...,\vec{g}'_n\right) \nonumber
\end{eqnarray}
where we have highlighted the fact that the GFT interaction kernels are nothing else than the matrix elements of the canonical projector operator in the basis of products of single-vertex states, and a mass term incorporates the chemical potential.
Notice that this also means that the spin foam vertex amplitudes, which are nothing else than the GFT interaction kernels (as we will discuss shortly), have to be understood as encoding the matrix elements of the projector operator of the canonical LQG theory, and not directly those of the Hamiltonian constraint operator\cite{AlesciNouiSardelli, thomasantonia}.

As we stressed, the above is quite heuristic. The GFT path integral and the quantum statistical partition function of the canonical LQG theory have to be properly defined, i.e. one has to control the whole quantum dynamics of the theory. The advantage of the GFT reformulation of the canonical LQG dynamics is exactly that, {\it starting} from the GFT partition function in terms of the classical GFT action, it allows to use a host of QFT tools, most notably renormalisation, to define properly the quantum dynamics, as we will discuss. In fact, a few general facts are however clear from the above correspondence. The GFT formulation of a LQG dynamics stands with respect to it as the field theory formulation of the dynamics of a many-body system with respect to its 1st quantised formulation in terms of a Schroedinger equation for many-body wave functions. The former, one expects, is the best set-up to tackle issues involving large numbers of degrees of freedom of the same system. In the same vein, the full quantum dynamics can obviously be studied perturbatively around the Fock vacuum (thus in terms of the spin foam expansion of the n-point functions, as we will see next), but this, as in standard QFT, is expected to be a good approximation of the full dynamics only for processes involving very few quantum geometric degrees of freedom of the kinematical type, i.e. only for physical situations in which the physical vacuum of interest is well approximated by the perturbative Fock vacuum \cite{GFTfluid, GFTlorenzo,VincentTensor}. Finally, the canonical projector equation for a Hamiltonian constraint operator, has to be looked for in the sub-sector of the full GFT quantum dynamics corresponding to a restriction to the micro canonical ensemble, possibly corresponding to a \lq tree-level\rq ~restriction of the GFT partition function \cite{GFT}.  

\

\noindent {\bf GFT dynamics from spin foams/lattice gravity path integrals -} The above strategy for the definition of GFT models, starting directly from a canonical quantum dynamics, has not been followed until now. The main strategy that has been followed starts from the definition of spin foam amplitudes, encoding them in a GFT model. This has been mainly done in a simplicial context \cite{GFT} (but see \cite{GFTallLQG}). 
 
Working in the simplicial complex means that one chooses $d$ equal to the would-be space-time dimension, interprets the GFT fields (i.e. the quanta they create/annihilate) as $(d-1)$-simplices (quantum tetrahedra in $d=4$), with the arguments of the GFT fields attached to their $(d-2)$-faces. Next, one usually restricts possible interactions to a single one, describing $d+1$ such simplices glued pairwise across their faces to form (the boundary of) a d-simplex. The kinetic term describes the gluing of two such d-simplices across a shared $(d-1)$-simplex (the GFT quantum being propagated from one interaction vertex to the next). With this combinatorics, one has a general action
\begin{equation}
S_{GFT}= \int [dg_i][dg'_i]\,\varphi^*(g_i)\,\mathcal{K}\left( g_i, g'_i\right)\,\varphi(g'_i) \,+\, \frac{\lambda}{(d+1)!}\int [dg_{ij}]\, \varphi(g_{1j})....\varphi(g_{(d+1)j})\,\mathcal{V}\left(g_{ij}\right) + c.c. \quad \nonumber.
\end{equation}
The GFT Feynman diagrams are then 2-complexes dual to (the 2-skeleta of) simplicial complexes. The GFT perturbative expansion will then give a prescription for summing over such complexes, weighted by quantum amplitudes. The 2-complex corresponding to each Feynman diagram is a collection of vertices connected by links, in turn bounding 2-cells. Each GFT interaction kernel assigns an amplitude to each vertex of such 2-complex, while the GFT propagator gives an amplitude to each link of the same, and encodes a prescription for connecting the variables entering the vertex amplitude to the ones appearing in the amplitude of a neighbouring vertex. The complete Feynman amplitude for the 2-complex is obtained by summing over common variables in vertex amplitudes and propagators. Now, the vertex amplitude $\mathcal{V}$ corresponding to a given GFT model is nothing else than the spin foam vertex amplitude characterising a given spin foam model, and the spin foam model is itself fully specified by the whole GFT Feynman amplitude. Notice that, just like the spin foam amplitudes can be recast in different forms, including trivialising the vertex amplitudes and including all the non-trivial ingredients inside edge amplitudes, there is a certain freedom in a GFT model to redefine kinetic and interaction terms, while maintaining the same expression for the Feynman amplitudes obtained by convoluting them. These amplitudes can be written, just as the GFT field itself and the GFT action, in different variables: group elements, Lie algebra elements or group representations. In terms of group elements, the (spin foam) amplitudes take the form of lattice gauge theories \cite{SF, P-R, mikecarlo, IoRuth, BiancaRenorm}, the lattice being the 2-complex. In Lie algebra variables, the amplitudes take the form of discrete gravity path integrals on the same lattice (or on the dual triangulation) \cite{flux, BO}. In group representations, the amplitudes take the standard spin foam form \cite{SF}. This correspondence between spin foam models and GFTs is one to one and fully generic \cite{mikecarlo,GFT}: for any given spin foam model, there is a GFT model such that the spin foam amplitudes are reproduced as Feynman amplitudes, and any GFT model defines a spin foam model in its perturbative expansion. Thus any procedure for defining a spin foam model gives automatically a definition of a GFT model. 

In the simplicial context, spin foam constructions in 4d have followed two strategies \cite{SF,GFT}. The first is: given a simplicial complex and a choice of classical phase space being the cotangent bundle of $SL(2,\mathbb{C})$ or $Spin(4)$ for each triangle, thus classical variables being Lie algebra elements, interpreted as bivectors associated to the triangle, and group elements (a discrete connection) associated to a dual link, impose appropriate conditions on such variables at the level of quantum states, such that only proper quantum simplicial geometries are summed over in the amplitudes. This can be called the state sum strategy. The second is: start from a spin foam model for topological BF theory, and define a 4d gravity model by imposing at the level of the amplitudes the discretised version of the simplicity constraints that turn the BF dynamics into the gravitational one in the continuum \cite{SF,Plebanski}. Given that such topological spin foam models can be equivalently written as discrete path integrals for BF theory, the constrained spin foam amplitudes become equivalently discrete path integrals for Plebanski gravity \cite{BO}. Moreover, the simplicity constraints amount also to a restriction in the decomposition of the representations of $SL(2,\mathbb{C})$ or $Spin(4)$ into representations of the diagonal $SU(2)$ subgroup, or, viceversa, an embedding of the latter into the former, allowing a link with canonical LQG. The crucial issue becomes of course the correct discretisation and quantum implementation of the simplicity constraints, which is where specific models differ. In any case, both general strategies coincide in their resulting spin foam amplitudes, due to the fact that the initial phase space variables assigned to tetrahedra in the state sum approach are exactly the phase space variables of discretised BF theory and that the simplicity constraints are indeed geometricity conditions for such tetrahedra. Both strategies can be followed directly at the GFT level as well, by imposing constraints on the GFT fields, representing, we recall, quantum tetrahedra.  

To illustrate the resulting GFT models, we give two examples in the Riemannian setting including the Immirzi parameter $\gamma$ (in the Lorentzian setting less is known and at present we only have one model, with minor variations: the EPRL model \cite{EPRL} or the BC model \cite{BCrevised}, depending on whether the Immirzi parameter is included or not). The same models can be straightforwardly extended to combinatorial settings more general than the simplicial one \cite{KKL, GFTallLQG}, but the geometric features of such generalised constructions are not yet fully understood. The first example is the version of the EPRL model \cite{EPRL,SF}, resulting from a specific choice of imposing the constraints in the representation variables, and only at the level of the GFT kinetic term:
\begin{equation}
S^{EPRL}_{GFT}\,=\, \int [dg_i]\,[dg'_i]\,\varphi^*(g_i)\,C^{-1}\left( g_i, g'_i\right)\,\varphi(g'_i) \,+\, \frac{\lambda}{5!}\int [dg_{ij}]\, \varphi(g_{1j})....\varphi(g_{5j})\,\prod_{i\neq j, i,j=1}^{5} \delta(g_{ij},g_{ji}) + c.c. \nonumber
\end{equation}
where the interaction term is the same as in BF models and simply identifies the $Spin(4)$ group arguments of the GFT fields according to the combinatorics of faces in a  4-simplex, while the kinetic term is determined by:
\begin{eqnarray}
&{}&C_{EPRL}\left( g_i, g'_i\right) \, =\, \sum_{j^+_i, j^-_i,J_i \in \mathbb{N}/2} \left( \prod_{i=1}^4 d_{j^+_i}d_{j^-_i}d_{J_i}\, \delta_{|1-\gamma|j^+_i,(1+\gamma)j_i^-}\,\delta_{J_i, j_i^+ + j_i^-}\right)\int dh_{\pm}dh'_{\pm} \int \prod_i du_i \nonumber\\ &{}&\,\left[ \prod_{i=1}^4 \chi^{j_i^+}\left( g_i^+ h_+ u_i (h_+')^{-1} (g_i^{'+})^{-1}\right)\,\chi^{j_i^-}\left( g_i^- h_- u_i (h_-')^{-1} (g_i^{'-})^{-1}\right)\chi^{J_i}\left( u_i\right)\right]\; \nonumber ,
\end{eqnarray}
where all the integrals are over $SU(2)$, the $Spin(4)$ group elements are decomposed into their selfdual/anti-selfdual components, $(j_i^+,j_i^-)$ label irreducible unitary representations of $Spin(4)$, while $J_i$ label irreps of the diagonal $SU(2)$ subgroup, and $\chi$ are the representation characters; here $\gamma$ has to be a rational number.
A second example is the version of the BO model \cite{BO} resulting from a specific choice of imposing the constraints in Lie algebra (flux) variables, again only in the kinetic term:
\begin{eqnarray}
S^{BO}_{GFT}\,&=&\, \int [dg_i]\,[dg'_i] dk dk'\,\varphi^*(g_i; k)\,C^{-1}\left( g_i, k; g'_i, k'\right)\,\varphi(g'_i; k') \,\nonumber \\&+&\, \frac{\lambda}{5!}\int [dg_{ij}]\,[dk_j] \varphi(g_{1j};k_1)....\varphi(g_{5j};k_5)\,\prod_{i\neq j, i,j=1}^{5} \delta(g_{ij},g_{ji}) + c.c. \nonumber
\end{eqnarray}
with the same BF interaction term, and an additional set of variables $k_i\in S^3\simeq SU(2)$, interpreted as unit normals to the tetrahedra, introduced to ensure the covariant imposition of the constraints, while the kinetic term is determined by:
\begin{eqnarray}
&{}&C_{BO}\left( g_i, k; g'_i, k'\right)\,=\, \int [dx_i][dy_i] \int dh^{\pm}dh^{'\pm} \, \delta\left( k' k^{-1}\right) \qquad \nonumber \\ &{}&\prod_i\left[  E_{g^+_i h^+}(x_i^+)E_{g^-_i h^-}(x_i^-) \star \delta_{-k x_i^- k^{-1}} (\beta x_i^+) \star \delta_{- x^+_i}(y_i^+) \delta_{- x^-_i}(y_i^-) \star E_{g^{'+}_i h^{'+}}(y_i^+)E_{g^{'-}_i h^{'-}}(y_i^-)\right] \nonumber
\end{eqnarray}
where $\star$ is the $\star$-product entering the definition of the Lie algebra representation in terms of variables $X_i = (x^+_i, x_i^-)\in \mathfrak{so}(4)$ and encoding the chosen quantisation map for flux operators, $E_g(x)$ are the corresponding non-commutative plane waves, satisfying $E_g(x)\star E_h(x) = E_{gh}(x)$, and $\delta_{-y}(x) = \int du E_u(y)\,E_u(x)$ is the corresponding non-commutative delta function \cite{flux, QuantFlux, BO}. This second model can easily be written down in group variables only (carrying out the Lie algebra integrations above).

As said, these choices of GFT action give rise to two different spin foam models, as the corresponding Feynman amplitudes are different. The general properties of the GFT definition of the spin foam models are the same, though: the spin foam amplitudes associated to a given spin foam 2-complex (or dual triangulation) are completed by a sum over complexes of any topology, generated perturbatively with the canonical combinatorial weights of a field theory, depending on the coupling constants and on the symmetry group of each complex, as in ~\ref{GFTPartFunc}. They can be equivalently expressed in group variables (where they take the form of lattice gauge theories), Lie algebra variables (where they take the form of simplicial gravity path integrals) or group representations. The resulting definition of the dynamics is covariant in the sense that it does not require a canonical 3+1 splitting, and it incorporates topology change in a very natural way. It is a form of discrete 3rd quantisation of gravity \cite{GFT3rd}, and a peculiar type of discrete realisation of gravitational path integral, combining both sum over lattices as in dynamical triangulations \cite{DT}, and a sum over discrete geometric data for each given triangulation, like in quantum Regge calculus \cite{RuthRegge}. It also gives a QFT generating functional for the same discrete gravitational path integral. Plus, combined with the previously discussed heuristic derivation of a GFT model directly from the canonical dynamics, the GFT encoding of spin foam models allows to link them directly with the canonical theory. 

The \lq spin foam\rq strategy is quite satisfactory from the point of view of encoding discrete geometry and making contact with discretised gravity. However, it leaves the definition of the {\it theory space}, e.g. the set of possible interactions that can/should be included in the theory, rather ambiguous. From the point of view of discrete geometry (and also of the canonical LQG theory), given the same quantum geometric data and a prescription for imposing simplicity constraints, one could consider adding more interaction terms with different combinatorics, which however may also require additional constraints to maintain geometricity of the complexes thus generated. Not to mention that the same strategy of constraining BF theory could be extended to formulations of gravity more general than the Holst-Palatini one. One possible attitude toward these issues is to rely on the fundamental nature of simplicial geometry and on the renormalisation group. One could argue that simplicial structures can be considered the most basic type of lattices on which to discretize geometry, and that one has to simply start from the simplest GFT action ensuring geometricity of the simplicial structures it generates, and then run the quantum dynamics and the renormalizationg group to generate all possible interactions compatible with that at different scales, the only constraint being renormalizability of the resulting model. Still, one may want to have a more principled definition of the GFT theory space \cite{vincentTS}, resting on basic assumptions, some GFT counterpart of QFT axiomatics. 

\

\noindent {\bf GFT dynamics from tensorial axiomatics -} This leads to a third strategy for the construction of GFT models. It partly stems from the search for a good notion of locality for GFTs by extending the basic features of matrix models for 2d quantum gravity \cite{matrix} to higher dimensions, stressing the interpretation of GFTs as (richer) tensor models \cite{tensor}. First, to be able to interpret the GFT field as a proper tensor, one has to define a transformation under a unitary group $U^{\times d}$, like: $\varphi(g_1,..,g_d) \rightarrow \int [dg_i] U(g'_1, g_1)\cdots U(g'_d, g_d)\, \varphi(g_1,...,g_d)$, and $\varphi^*(g_1,..,g_d) \rightarrow \int [dg_i] U^*(g_1, g'_1)\cdots U^*(g_d, g'_d)\, \varphi^*(g_1,...,g_d)$ for $\int dg U(g'_i,g_i)^* U(g_i, \tilde{g}_i) = \delta(g'_i, \tilde{g}_i)$, which in turn requires the $d$ arguments of the GFT field to be {\it labelled}. Given the tensorial transformation property, one can define {\it tensor invariant} interactions corresponding to invariant convolutions of polynomials of GFT fields. Such invariants have a nice graphical representation: they are labelled by {\it coloured d-graphs} $\mathcal{B}$ constructed as follows: for each GFT field (resp. its complex conjugate) draw a white (resp. black) node with $d$ outgoing links each labelled by $d$ different colours, then connect all links with one another following the two conditions that a white (resp. black) node can only be connected to a black (resp. white) node and that only links with the same colour can be connected. The resulting functionals $I_b$, with $b\in \mathcal{B}$, are invariant under $U^{\times d}$ (for real fields, one has a similar definition involving orthogonal transformations). Moreover, this is the tensor invariance property generalising the invariance of matrix models ($d=2$), whose interactions can only be traces of matrix polynomials. In the matrix case, this invariance indeed replaces the notion of locality of usual QFT. This suggests a GFT theory space with arbitrary tensor invariant interactions and a kinetic term which may instead break the invariance (like one uses non-local kinetic terms in usual QFT):
\begin{equation}
S_{GFT}\,=\, \int [dg_i]\,[dg'_i]\,\varphi^*(g_i)\,\mathcal{K}\left( g_i, g'_i\right)\,\varphi(g'_i) \,+\,  \sum_{b \in \mathcal{B}} t_b I_b (\varphi , \varphi^*)
\end{equation}
where $t_b$ are the coupling constants corresponding to the various tensor invariant interactions.
The use of tensor invariant interactions has a very nice property at the topological level. The Feynman diagrams of such GFTs can be represented as $(d+1)$-coloured graphs, in which the $d$-coloured graphs representing interactions are of course associated to vertices of the diagrams and an extra colour is associated to the lines of propagation of the GFT field. It is an important result in crystallization theory \cite{crystal} that $(d+1)$-coloured graphs are in one to one correspondence with simplicial pseudo-manifolds, i.e. manifolds with at most conical singularities. Moreover, when complex fields are used the graphs are bipartite and these are in correspondence with {\it orientable} pseudo-manifolds \cite{tensor}. These topological properties give an additional motivation to follow this third strategy, while one has also to notice that, up to now, this strategy has offered less indications in the choice of field content and of kinetic/interaction kernels. 

The same topological considerations show also an interesting link between models based on simplicial interactions (one may say: with \lq simpliciality\rq ~replacing locality) and tensor invariant models. Starting from simplicial interactions, one notices that the data in a GFT Feynman diagram are sufficient to determine a 2-complex dual to a simplicial complex, but not to specify the full homology of the same complex, in particular the $k$-cells for $k< d-2$. This is a problem, in particular if one wants to understand the symmetry properties of the lattice gauge theories, discrete gravity path integrals and spin foam models associated to the same Feynman diagrams \cite{GFTdiffeo, MatteoValentinCellSF}. A solution to this problem \cite{tensor, colouredGFT} is again to add {\it colours} to simplicial GFTs. Instead of working with a single d-valent GFT field, one uses $d+1$ such fields $\varphi_a$, and their complex conjugates, labelled by a colour index $a=1,..,d+1$, with a kinetic term for each of them, and a single simplicial interaction involving a convolution of all of them, one for each $(d-1)$-simplex in a d-simplex, plus the complex conjugate of the same interaction. Feynman diagrams will again be $(d+1)$-coloured graphs, representing the 1-skeleta of simplicial pseudo-manifolds. Moreover \cite{uncoloring}, starting from such coloured simplicial models and integrating out all but one of the GFT fields (these functional integrations are gaussian), one is left with a theory of a single GFT field with interactions indexed by coloured d-graphs $\mathcal{B}$. This intriguing connection deserves to be further explored.

\

\noindent {\bf GFT symmetries -} Still, much remains to be understood about basic principles determining the GFT theory space. In particular, one would like to understand more about GFT symmetries. For GFTs aiming at quantum gravity, one would like to understand better especially the role of diffeomorphisms. Here as well there are three main lines of attack, obviously not exclusive of one another, again depending on whether one approaches GFTs from the canonical LQG side, from the discrete gravity or spin foam side, or from the tensor models side. Coming from the canonical theory, one would make sure that some Hamiltonian (and spatial diffeo) constraint operator is encoded in the quantum GFT dynamics by construction, and look for the sector in which only quantum states satisfying the corresponding operator equations are relevant. This route has not been followed up to now, if not indirectly via the spin foam route. Following the spin foam or discrete gravity route, one would first look for symmetries satisfied by the GFT Feynman amplitudes, some discrete counterpart of diffeo symmetry in the continuum, and then look for the field-theoretic definition of the same at the GFT level. Simplicial diffeos have been studied in both discrete gravity \cite{biancaDiff} and topological spin foam models \cite{LaurentDiffeo}, and in fact, their GFT counterpart has been identified \cite{GFTdiffeo}. Interestingly, on the one hand this turns out to be a {\it global} quantum group symmetry, at the GFT level, on the other hand it leads naturally to the discrete WdW-like equation and recursion relations seen also at the spin foam level \cite{SFrecursion}. These simplicial diffeos are found in topological models, but are broken as soon as simplicity constraints are introduced, leaving open the issue of recovering them in a suitable approximation of gravitational GFT models in 4d. A third complementary strategy follows again the insights of matrix models. There \cite{matrix}, the Virasoro algebra is found from a set of constraints satisfied by the partition function, in turn following from the Schwinger-Dyson equations of the n-point functions associated to loop observables. The latter can also be directly related to the WdW equation of 2d Liouville gravity corresponding to the continuum limit of the same matrix models. An analogue result is found in simple tensor models \cite{tensor,virasoro}, where a generalisation of the Witt algebra is found by a clever rewriting of the Schwinger-Dyson equations. However, not only we still miss the link between this symmetry algebra and higher-dimensional diffeomorphisms, but we need to generalise these results to more involved GFTs with interesting quantum geometric data. In general, the treatment of GFT symmetries as in standard QFTs needs to be better developed \cite{JosephClassicalGFT}: e.g. concerning a GFT analogue of Noether theorem and classical conserved currents, a detailed understanding of quantum Ward identities for n-point functions and of quantum symmetry breaking. 

\section{The continuum limit of quantum geometry in GFT}
Despite GFTs being formally continuum QFTs, they give a picture of space-time as fundamentally discrete, in the same sense in which QFT has a fundamental discreteness in it: the fundamental degrees of freedom that (should) make up space-time are GFT quanta (simplices or spin nets vertices), as brought to the forefront by the 2nd quantised formulation, and their dynamical processes are the GFT Feynman diagrams, themselves cellular complexes replacing a smooth space-time manifold. Further, if the group manifold, on which the GFT field is defined, is chosen to be $SU(2)$ or another compact group, one obtains an additional discreteness of quantum geometric spectra, such as areas and volumes, as in standard LQG.

Still, our conventional description of gravity and space-time is in terms of a continuum field theory, General Relativity, based on a smooth metric field on a smooth manifold (it is in this sense that the Regge geometries found in GFT Feynman amplitudes, which are continuous but piecewise flat, thus singular, are not enough). To reconcile this description with the one provided by GFTs is the problem of extracting an effective continuum description for the GFT degrees of freedom. 

\

\noindent {\bf Understanding the nature of the problem -} Before we move on to discuss this area of GFT research, we would like to emphasize the difference with the problem of the classical limit. 
By continuum approximation in this context we mean mainly a regime in which a large number of fundamental degrees of freedom, i.e. GFT quanta, are effectively accounted for in terms of some collective variables (possibly with additional physical constraints). By classical approximation we mean the regime in which the quantum nature of the same degrees of freedom (few or many) can be neglected. The two approximations or limits do not need to commute at all, and it is then very important to explore both of them independently. In particular, it could well be that the quantum nature of the fundamental degrees of freedom is actually {\it needed} to achieve the correct continuum effective physics, i.e. the classical approximation may need to be taken {\it after} the continuum approximation. From this perspective, the usual semiclassical approximations in spin foam models \cite{SF} (e.g. the asymptotic analysis, showing dominance of Regge geometries for each spin foam 2-complex) and canonical LQG (e.g. coherent states peaking on discrete classical geometric data for any given spin network graph \cite{coherent}), do not ensure the existence of the right continuum limit, and it is not even clear that such continuum limit has to involve only these approximately classical configurations. 
From this perspective, the continuum limit in GFTs is rather to be understood in analogy with the thermodynamical limit in usual QFTs (e.g. in condensed matter) as the regime of (formally) infinite fundamental atoms and of various cut-offs (both UV and IR) being (formally) removed. 

The problem of the continuum in GFT, i.e. the problem of controlling the quantum dynamics of the theory when more and more interacting degrees of freedom are taken into account, can then be articulated in four main related aspects. One is the perturbative renormalizability of GFT models. The second is the unraveling of the phase structure of the same models. The third is their non-perturbative, constructive definition. The fourth is the extraction of effective continuum physics from them, in the appropriate regime. We will discuss the last, more physical issue in the next section. Here we will deal with the first three, more formal issues. 

In order to appreciate further their relevance, let us first discuss what role these aspects of GFTs play from the point of view of canonical LQG and of spin foam models. 
Both GFT renormalizability and constructibility are simply equivalent to the requirement of GFT models be (non-perturbatively) well-defined. As such, they are a re-phrasing of the requirement of having a well-defined dynamical constraint operator. Related to the same definition of the canonical constraint operator, one should worry about quantisation ambiguities. These have been related to the issue of renormalizability in perturbative QG \cite{AlexAmbig}, and its GFT counterpart is of course again GFT renormalizability. Even if the theory was entirely well-defined mathematically, an important physical question would remain: what is the flow of effective quantum dynamics across scales? what is the continuum phase structure? These issues are crucial both in the canonical and in the covariant setting. They can be tackled very effectively in the GFT formalism, thanks to the QFT setting. It is not so clear that they can be tackled as effectively remaining in the canonical operator formalism, and the use of QFT methods rather than canonical operator methods in many-body quantum theory would suggest otherwise. As for the spin foam context, if the spin foam expansion is understood as arising from expansion of canonical evolution operator (but this picture is not so consonant with the relation between canonical dynamics and GFT one as discussed above), the problem is to make sense of this expansion, i.e. to define the canonical evolution operator rigorously, and becomes again the issue of finding a constructive definition of the GFT partition function. If spin foams are understood as lattice gravity path integrals, the problem of the continuum translates into the the issue of refining/coarse graining them, and identify the flow of effective dynamics and its fixed points \cite{BiancaRenorm}. That is, the same problems that GFT recast as in QFT language by turning the same lattice gravity path integrals into Feynman amplitudes. Regardless of the precise point of view one takes about what spin foam models are, one thing is clear: there is simply no spin foam theory without a prescription of how to deal with spin foam amplitudes on different complexes. In other words, a complete definition of a spin foam theory is not simply the set of all possible spin foam amplitudes for all possible spin foam complexes, and the relative physics is not to be extracted simply by choosing some given spin foam complex, adapted to the situation at hand, on every occasion. Rather, a complete definition is given by: this set of complexes {\it plus} a precise organisation principle, a procedure to relate the amplitudes for different complexes. Because the set of all complexes has to include the ones made of infinite numbers of cells/links/faces, if theory is to have an infinite number of degrees of freedom, this becomes the problem of defining the continuum limit. A formal definition of such organisation principle specifies a formal definition of the theory, then to be made rigorous. This principle could be a refinement limit and a coarse graining procedure (see \cite{BiancaRenorm}) or a prescription for summing over complexes. The latter could be chosen arbitrarily, but this introduces a further huge ambiguity. Rather, it should be selected by an appropriate set of principles (maybe motivated by the canonical perspective \cite{thomasantonia}). The GFT formalism provide exactly such prescription, a natural and clear organisation principle for spin foam amplitudes, and for defining their continuum limit.

\

\noindent {\bf A survey of current research -} In pursuing the issue of renormalizability and constructibility of GFTs, the input and recent results from tensor models have proven (and will certainly prove in the future) crucial. In fact one main obstacle towards making sense of the continuum limit of GFTs is to control the sum over triangulations/spin foam 2-complexes they generate, which are much more intricate in their combinatorial topology, thus more difficult to handle, than standard QFT Feynman diagrams. 
On this progress has been made, thanks to developments in tensor models. The introduction of colours \cite{tensor, colouredGFT} gave control over the combinatorics and topology of GFT Feynman diagrams, and over their sum. One important result has been the understanding of the large-N limit of tensor models (where $N$ is the size of the tensors) and topological GFTs (where $N$ is the cut-off in representations) \cite{large-N}, which are the starting point for the construction of most 4d gravity models. It has been shown that the leading order in $N$ corresponds to {\it melonic} diagrams \cite{critical, uncoloring, tensor}, which are dual to triangulations of spheres with a peculiar combinatorial structure, maximising the number of faces for given number of vertices of the 2-complex; thus, they are expected to dominate the expansion of any model whose amplitudes scale with the number of faces. Important universality results have also been obtained \cite{universality}. 

The ability to control the types of cellular complexes appearing in the GFT perturbative sum is in obviously crucial for GFT renormalization, which started being tackled in \cite{GFTrenorm1} and is, since then, a very active research area \cite{GFTrenorm}. Next to the mentioned combinatorial aspects, this involves a deeper understanding of the scaling of amplitudes in the large-N limit.  Indeed, the large-N regime is the regime of many GFT degrees of freedom (for given combinatorics). From this point of view it represent the analogue of the UV regime in usual QFTs. Conversely, the low-N regime would be the analogue of the IR regime. Consistently with this picture, it is in this large-N regime that divergences in spin foam amplitudes are usually found. However, the geometric interpretation associated to it is not totally clear. Large spins in $SU(2)$ spin foam models correspond to large areas and volumes for the discrete structures they are based on, consistently with the quantum geometry of canonical LQG. From the point of view of simplicial geometry, then, the geometric notion of UV/IR is rather the opposite of the formal QFT one. Still, much caution should be exercised in interpreting in geometric terms the algebraic data and discrete structures of GFTs (and spin foams), before a proper continuum limit is established and we control how such data and structures map to effective continuum geometric ones. These considerations affect work on GFT renormalisation only to a limited extent, as one can proceed guided only by the mathematical behaviour of GFT amplitudes. Their scaling with $N$ is the first thing that has been studied, with focus on (coloured) topological models, in particular showing the suppression of singular topologies \cite{pseudo}.
Still in the context of topological GFTs, remarkable calculations of radiative corrections were performed \cite{valentinjoseph}, and one interesting implication was that, in order to achieve renormalizability, these models need to be augmented by a kinetic term given by the Laplace-Beltrami operator on the group manifold
$$
\int [dg_i]\,\varphi^*(g_i)\,\sum_{i=1}^d\Delta_{G_i}\,\varphi(g_i)\qquad.
$$
 While its quantum geometric meaning is still unclear, this is indeed a natural choice of kinetic term, and it has been later shown to make topological models superrenormalizable (at least in the abelian case) \cite{GFTrenorm}. Similar, and much more challenging computations have to be performed in a more systematic way in the context of 4d gravity models. While the Barrett-Crane model can be shown (or argued) to be super-renormalizable as well, without modifications, at least for special choices of edge amplitudes \cite{P-R,baez,BCrevised}, little is known about models incorporating the Immirzi parameter, like the EPRL and BO models. The simplest radiative corrections of the EPRL model have been investigated in \cite{EPRL} and, much more thoroughly, showing the many interesting intricacies of the quantum amplitudes, in \cite{aldo}, and are similarly dependent on the details of edge weights.

While these calculations are technically impressive and important, one would like a more systematic analysis of perturbative renormalizability, which in turn requires more control on the theory space. Indeed, up to now this type of systematic analysis has been carried out only in simplified models characterised by tensor invariant interactions and the mentioned Laplacian-type kinetic term (this class of models has been dubbed {\it tensorial GFTs}). While the use of tensor invariant interactions provides a notion of locality, the presence of a Laplacian term endows the models with a proper notion of {\it scale} (indexed by its eigenvalues, i.e. by group representations, as expected from LQG and spin foam models). This allows to apply to the more involved GFT context the rigorous multi scale renormalization analysis of conventional QFTs \cite{vincentBook}. Results have been piling up rapidly. After the identification of several renormalizable models in both 3 and 4 dimensions \cite{GFTrenorm}, the investigation of models incorporating the gauge invariance conditions characterising topological GFT models (and crucial also for gravitational models in 4d) has started. Renormalizable models were identified, of the abelian type, in several dimensions \cite{COR1,GFTrenorm}. Lately, a non-abelian GFT model with gauge invariance, with interactions up to order six, extending the Boulatov model for topological BF to include a Laplacian term, was also shown to be renormalizable\cite{COR2}. 
This type of analysis is highly non-trivial, due again to the intricacies of cellular complexes, requiring an extension or adaptation of several notions and tools from standard renormalisation theory: the notion of Wick ordering, the notion of connectedness and of 1-particle irreducibility, that of contraction of high subgraphs, as well as several tools from crystallisation theory such as the use of {\it dipole moves} to contract Feynman diagrams. 
The stage is now set for a similar systematic analysis of gravitational GFT models in 4d, incorporating also simplicity constraints. A renormalizability result in this context, for models of the EPRL or BO type, would be of paramount importance for the whole field of quantum gravity.
Renormalizability (or perturbative finiteness, which is nicer in some respect but more problematic in others, as it makes it harder to identify the relevant channels of interactions at different scales) would imply that the given GFT model is perturbatively well-defined, as a QFT. It would represent a {\it truly background independent and renormalizable quantum (field) theory of spacetime}. Given that how spin foam models arise in GFT, perturbative GFT renormalizability would also give meaning to the spin foam sum over complexes.

The renormalisation group also determines the flow of effective GFT dynamics across scales, and helps us to map out the phase structure of the theory. This is being done as well. Detailed calculations of the beta functions for various tensorial GFTs have been performed \cite{beta}, again culminating in the recent analysis of non-abelian models with gauge invariance \cite{sylvain}. While establishing general conclusions is very tricky\cite{sylvain}, the argument can be made for asymptotic freedom (or safety) being rather generic in GFTs. Asymptotic freedom would of course be welcome news for GFT models of quantum gravity because it would confirm their being well-defined as QFTs. It would also suggest the existence of phase transitions at some point during the flow towards small values of $N$, i.e. small values of group representations or Lie algebra elements (simplicial areas and volumes), due to a corresponding growth in the GFT coupling constants. 
One powerful tool to establish such flow and to study it, with the aim of mapping out the GFT phase diagram, is the so-called Functional Renormalization Group, central also in the Asymptotic Safety approach to quantum gravity \cite{AS}. An extension of the FRG to matrix models has been developed in \cite{AstridTim} and suggested to generalise to tensor models as well. The first definition and application of the FRG to tensorial GFTs has been indeed provided recently \cite{IoDarioJoseph}. Beside making available this powerful tool, it allowed to obtain interesting informations about fixed points and the phase diagram of simple models (without gauge invariance) at both large and small $N$, and the first indications of a phase transition between a phase characterised (in mean field) by a vanishing GFT field, and a \lq condensed\rq phase with non-vanishing expectation value of the same. The possible physical relevance of this scenario will be discussed in the next section.  

The analysis of phase transitions in GFTs is in its infancy. The only other result is the proof of existence of a phase transition for any topological (BF) model in any dimension, in the melonic sector \cite{MelonicTrans}. More is known in tensor models. Here the problem is tackled by explicit resummation of the partition function, in the appropriate regime, i.e. from a statistical rather than field-theoretic point of view. For both i.i.d. and dually weighted models, not only the critical behaviour of the melonic sector has been studied, with the explicit calculation of critical points and critical exponents \cite{critical, uncoloring}, but subdominant orders have also been resummed and a rigorous double scaling limit has been performed, again with explicit calculation of the critical behaviour \cite{tensorDouble}. Once more, work on these simpler models may pave the way for the analysis of GFTs for quantum gravity.

Results on GFTs should be compared with analogous results in canonical LQG and in the lattice gravity approach to spin foam models. In the canonical setting, these issues have not been much explored, though, and what we know at present is only that inequivalent kinematical representations of the holonomy-flux algebra exist, beyond the Ashtekar Lewandowski one. In particular, we have the Koslowski-Sahlmann vacuum \cite{HannoTim} with its non-degenerate intrinsic geometry and spread-out extrinsic curvature, and, in the simplicial context, the BF vacuum obtained imposing exact flatness of the connection, thus with spread-out intrinsic geometry \cite{BiancaMarc}. Also, work on GFT renormalisation should be more carefully compared with renormalisation of spin foam models treated as lattice gauge theories \cite{BiancaRenorm}. In fact, from the point of view of their Feynman amplitudes, the procedures of subtraction of subgraphs adopted to establish perturbative GFT renormalizability can be seen as lattice coarse graining procedure, thus reinterpreted from a different perspective.

To conclude, we mention results on constructive aspects of GFTs, aiming at a non-perturbative definition of GFTs: the perturbative series should not only be renormalizable but also \lq\lq summable" (it is not going to be convergent, in general). The summability in the melonic (large-N) sector has been already discussed, as well as the one involving subdominant orders, all involving complexes of trivial topology. Interesting results have been obtained also on the Borel summability of the whole GFT partition function for topological models \cite{Borel}. These results, although their physical meaning is yet to be explored, are remarkable because they amount to being able to sum over all cellular topologies generated by the GFT perturbative expansion, a remarkable feat previously achieved only in the much simpler matrix models. More remains to be done, and once more results on constructive tensor models \cite{ConstructTensor} will be an important reference.

\section{Extracting effective continuum physics from GFTs}
\noindent {\bf The problem of continuum and GFT phase transitions from a physical perspective -} The crucial problem for any quantum theory of gravity, which starts from a background independent formulation, is to recover classical GR on a smooth manifold in the continuum and classical limit. Many results have been obtained in the canonical setting as well as in spin foams. Being a reformulation of canonical LQG and a complete encoding of spin foams, many of the techniques and results obtained in these two contexts are immediate to import in GFT. Also, in models defined in the simplicial context,  standard results from Regge calculus and related formalisms can be imported as well. Among these many results, here has been important progress in the construction of semiclassical states on given lattices/graphs (e.g. coherent states \cite{coherent}), and results on the semiclassical approximation of spin foam amplitudes\cite{SF}, giving Regge geometries, which in turn we know give Einstein geometries in the continuum limit (refined lattices, small curvatures and small volumes). These are important assets, but the important points we have already mentioned in the previous section should not be forgotten: first, the classical approximation is independent and possibly secondary to the continuum one; second, simple graphs/lattices (few degrees of freedom) could be useful to capture some limited information of continuum geometry and physics, in very special regimes, but effective continuum physics requires the limit of formally infinite number of fundamental degrees of freedom (associated to discrete structures). 
This is also the reason why recent attempts to compute effective continuum physics using spin foam models as such are problematic, from the GFT point of view, be them gravitational perturbations on flat space \cite{graviton} or cosmological equations \cite{SFcosmology}. These spin foam calculations are indeed confined to a regime, associated to simple spin foam 2-complexes (the perturbative GFT regime at the lowest order), and to a class of states, simple spin network graphs (very few GFT quanta), very far from the non-perturbative, approximately continuum sector of the theory, and instead too close to the regime of fully degenerate geometry (GFT Fock vacuum). The insights they provide are certainly useful, but something different is needed. Let us clarify further.

The GFT formulation of canonical LQG and spin foam models plays  provides indeed tools to deal with many QG degrees of freedom, i.e. to control the superposition of states and amplitudes associated to different and refined lattices. Just like in condensed matter systems the QFT language is useful to control many-particle physics and to extract effective dynamics from it, similarly the GFT language in the context of LQG. Like QFTs in condensed matter, one can deal with the GFT dynamics perturbatively, that is in its spin foam expansion. However, this should be expected to be the right language as far as few degrees of freedom are involved, that is close to perturbative GFT vacuum. In turn this perturbative vacuum is physically a \lq no-space state\rq, a fully degenerate geometry and topology. In contrast, one would expect non-degenerate geometries and effective continuum physics to arise far from such vacuum state and rather close to some different non-perturbative vacuum (a different phase) of the theory. This expectation is consistent with the strict canonical LQG picture as well, where the kinematical Ashtekar-Lewandowski vacuum is  fully degenerate one from the point of view of geometry, and simple spin network excitations around it are not enough to generate a smooth geometry. There are two interpretations one could give to this \lq need for a new, geometric phase\rq. The first treats this as a purely formal requirement, as simply implying that the theory is only physical in one appropriate phase (i.e. for the appropriate range of parameters), but with no meaning attached to other phases of the theory or to the phase transitions between them. For the range of parameters corresponding to the \lq right\rq phase, one then needs to study physics of both few and many d.o.f.s, and continuum physics is to be looked for when (formally) infinite d.o.f.s are involved. This is of course a conceptually consistent picture and, for what we know, could be the right one (it is, for example, the picture behind most work on the dynamical triangulations approach to quantum gravity \cite{DT}). A second interpretation, however, could also be put forward. It supposes that the other phases of the system may have a physical existence as well, even though they happen not to describe our current geometric regime of physical universe. In this view, the transition between phases may also have physical meaning, that is it could correspond itself to a physical process of the quantum gravity building blocks of space-time. More specifically, one could suggest that the phase transition to geometric phase is what replaces big bang singularity and describes the \lq origin\rq of our physical universe. That is, the hypothesis is that of a cosmological interpretation for the GFT phase transition to a non-degenerate, non-perturbative geometric GFT vacuum. This hypothesis can be dubbed \lq geometrogenesis\rq \cite{emergenceQG}.
While the search for such phase transition at the more rigorous level continues as summarised in the previous section, it is worth exploring possible concrete scenarios for it in GFT, aiming for a more direct extraction of possible physical consequences, and better physical insights into the formalism. This also means exploring candidates for the new vacua. A further hypothesis can then be put forward: our geometric universe could be born from a {\it condensation} of quantum space atoms \cite{GFTfluid}. This would realise explicitly in the GFT setting the idea of space-time as a condensate and of an emergent universe, often suggested in the context of analogue gravity models \cite{analogue,hu}. 

\

\noindent {\bf Some recent results -} Thus one turns the attention to a special class of states within the GFT Hilbert space: GFT condensates \cite{GFC}, that is quantum states characterised by a macroscopic occupation number for some given quantum observable \cite{leggett}, and involving a superposition of an arbitrary number of GFT quanta. In the simplest case, these are states such that all the GFT quanta are in the same quantum state. One can show \cite{GFC}, using a variety of results on LQG states and discrete gravity, that such condensate states, provided geometricity conditions (e.g. simplicity constraints) are imposed in the  quantum dynamics, admit an interpretation as continuum homogeneous spaces, of the type used in cosmology. They remain quantum states of the full theory, but they are fully characterised by a collective wave function depending only on the quantum geometric data associated to homogeneous anisotropic geometries. This characterisation still leaves room for a variety of constructions \cite{GFC, GFClorenzo, HomoNew}, but the simplest condensate state one can write down (the GFT analogue of the Gross-Pitaevskii ansatz for BECs):
\begin{equation}
|\sigma\rangle := \mathcal{N}(\sigma) \exp\left(\hat\sigma\right)|0\rangle \quad\text{with}\quad \hat\sigma := \int (dg)^4\; \sigma(g_1,\ldots,g_4)\hat\varphi^{\dagger}(g_1,\ldots,g_4) 
\end{equation}
where $\sigma(k g_1 ,\ldots,k g_4 )=\sigma(g_1,\ldots,g_4)\, \forall k\in Spin(4) \text{or} SL(2,\mathbb{C})$, and $\mathcal{N}(\sigma)$ is a normalization factor, is already very interesting. This state is quite special, as it is a coherent state for the GFT field operator:  $\hat{\varphi}(g_{I})| \sigma \rangle ={\sigma}(g_{I})| \sigma \rangle$, and using it as a (coarse grained) proxy of the true vacuum state is a sort of mean field approximation. Inserting this ansatz for the vacuum state of the theory in the quantum dynamics, that is in the Schwinger-Dyson equations, and making some further approximations, one gets an effective equation for the collective wave function $\sigma$ \cite{GFC}:
\begin{equation}
\int (dg')^4\;\tilde{\mathcal{K}}(g_1,\ldots,g_4,g'_1,\ldots,g'_4)\sigma(g'_1,\ldots,g'_4)+\lambda\,\frac{\delta\tilde{\mathcal{V}}[\varphi,\varphi^*]}{\delta\bar\varphi(g_1,\ldots,g_4)}\Big|_{\varphi\rightarrow\sigma,\varphi^*\rightarrow\sigma^*}=0\,.
\label{simpleeq}
\end{equation}
 which is nothing else than the classical equation of motion of the initial GFT model, up to any additional approximation needed to ensure consistency of the approximations and interpretation used, and symbolised by the $\tilde{}$ on the dynamical kernels. Given the interpretation of the collective wave function $\sigma$ as a distribution over the space of continuum homogeneous geometries, this equation represents a non-linear extension of the usual WdW-like equation of quantum cosmology, in particular, given the fundamental variables chosen, of loop quantum cosmology \cite{LQC}. Extended quantum cosmology equations of this type have been previously suggested in \cite{martin,GFCold}. The crucial point is that such quantum cosmology equation is here {\it derived from the fundamental theory}, for a suitable class of states, and no minisuperspace reduction is carried out (it should be compared with other interesting approaches to the same issue, like \cite{alescicianfrani}); rather, (generalised) quantum cosmology emerges from the fundamental dynamics as a kind of hydrodynamics approximation. Indeed, this derivation is compatible with the general expectations we discussed concerning the continuum approximation of the theory and phase transitions. Moreover, the derivation is completely general (for 4d gravity models with geometricity conditions), and it applies for any choice of fundamental GFT (and thus LQG and spin foam) dynamics, encoded in the GFT kernels $\mathcal{K}$ and $\mathcal{V}$. A simple example worked out already in \cite{GFC}, for both Lorentzian and Riemannian gravity, also incorporating a massless scalar field, and in more detail in \cite{GFCsteffen,GFCgianluca}, corresponding to using the Laplacian operator as $\mathcal{K}$ and neglecting the GFT interaction, already shows how a semi-classical Friedmann equation could be obtained. Similar results are obtained \cite{GFC} for different choices of GFT condensate states.
 
This equation can also be written down in terms of expectation values of dynamical operators, in turn involving the collective operators corresponding to cosmological variables \cite{GFCnew}. The natural canonical pair of cosmological variables, in terms of which one should try to write any effective semiclassical cosmological dynamics in a GFT setting, is given by (the expectation value of) the total flux $\hat{b}^i_I = i\kappa\int (dg)^4\; \hat{\varphi}^{\dagger}(g_J)\frac{{\rm d}}{{\rm d}t}\hat{\varphi}\left(\exp\left(\tau^i_I\,t\right)g_J\right)\Big|_{t=0}$ ($\kappa$ is a combination of Planck's and Newton's constants), from which geometric quantities like macroscopic areas and volumes are extracted, and the \lq average holonomy\rq $\hat\Pi[g_I]^{{\rm av.}} = \langle \hat{\Pi}[g_I] \rangle / \langle \hat{N} \rangle$, which contains extrinsic curvature information, computed from the extensive \lq total holonomy\rq $\hat{\Pi}[g_I] = \int (dg)^4\; \vec\pi[g_I]\;\hat{\varphi}^{\dagger}(g_J)\hat{\varphi}(g_J)$, for a choice of coordinates on the group $g =  \sqrt{1-\vec{\pi}[g]^2}\,{\bf 1} - i\vec{\sigma}\cdot\vec\pi[g]$, and the number operator $\hat{N}$. If these are the (expectation values of the) operators coming directly from the full theory, one sees immediately that the theory incorporates crucial quantum corrections to the usual classical variables. In fact, the total flux is, like its microscopic counterpart, non-commutative, while the standard cosmological variables correspond to its commutative limit $\hat{f}^i_I = i\kappa\int (dg)^4\; \hat{\varphi}^{\dagger}(\pi[g_J])\frac{\partial}{\partial\pi_i^I}\hat{\varphi}(\pi[g_J])$. Likewise, from the average holonomy one can define a macroscopic connection variable: $\vec\omega:=-\frac{\langle N\rangle^{1/3}\langle\vec{\Pi}\rangle}{|\langle\vec{\Pi}\rangle|}\,\arcsin\frac{|\langle\vec{\Pi}\rangle|}{\langle N\rangle}$. This second fact is important in several ways. To start with, one sees that the connection enters the effective cosmological dynamics necessarily with quantum holonomy corrections encoded in the sine function, as in loop quantum cosmology. Next, and most important, one see that the effective holonomy carries a dependence on the number of fundamental cells forming the universe. Again, this is like the lattice-refinement scheme in loop quantum cosmology, but here not only the dependence on $N$ is derived and not assumed, but this $N$ is a new 2nd quantised quantum observable of the theory. It enters necessarily both the kinematics and the effective cosmological dynamics \cite{GFCnew}. Finally, the same effective dynamics will relate the expectation value of $N$ and that of $a$, the scale factor, in such a way that cosmological holonomies end up depending on the scale factor as well, like in the so-called $\bar{\mu}$-scheme of LQC. For more details, we refer to the literature \cite{GFC,GFClorenzo,GFCsteffen,GFCgianluca,GFCnew}. It is clear, however, that it opens up a host of promising paths to explore, concerning, for example: the detailed analysis of effective cosmology coming out of various fundamental GFT models (e.g. EPRL or BO), the study of improved ansatz for condensate states (with a more detailed encoding of topology, involving better correlations between GFT quanta, etc); most important, a new way to study cosmological perturbations, understood as fluctuations above the GFT condensate, for which the derivation of an effective field theory picture is the most pressing issue. Beside the many intriguing conceptual aspects of this new picture of cosmology emerging from the full QG theory, this seems also a promising avenue towards extracting testable predictions from the fundamental theory and placing it in direct contact with observations.

We close by mentioning other results aiming at the extraction of effective continuum physics from GFTs, and working also in the spirit of mean field theory. Their difference in perspective notwithstanding, these were all obtained for GFT perturbations around (approximate) solutions of the classical GFT equations. They range from the derivation of an effective non-graph changing Hamiltonian constraint for spin networks starting from a GFT of topological type \cite{effHamilt}, to the extraction of classical equations for geometric phase space variables from the same GFT equations, using LQG coherent states as ansatz for the approximate solutions \cite{ioLorenzo}. This last work is close in spirit to the extraction of cosmology from GFT condensates discussed above. Also, there has been interesting work focusing on the effective dynamics of perturbations around classical GFT solutions, and showing how they can be recast in the form of non-commutative scalar field theories on a non-commutative flat space \cite{emergentmatter}, at least for specific models and specific choices of perturbations. This is an interesting result, which should now be re-analysed in the context of GFT condensate cosmology to be better understood and to acquire a more solid basis.

\section{Conclusions}
We have provided a quick survey of the GFT picture of quantum geometry and of its dynamics, and of some of the results in this area. We have tried to clarify the links (and differences) with canonical loop quantum gravity as well as with covariant spin foam models, emphasizing that GFTs offer a new elegant description of a quantum gravity theory based on spin networks, and at the same time a completion of the spin foam dynamics. This quantum field theory framework, in particular,  offers new tools that may prove crucial for addressing the issue of the continuum approximation and for the extraction of effective continuum dynamics from the full theory. In the long run, the goals of this approach are clear. We aim for: 1) a reliable (class of) model(s) for 4d Lorentzian quantum gravity with matter, with a nice and well-understood encoding of quantum geometry at the microscopic scales; 2) a proof that the same (class of) model(s) is perturbative renormalizable and possibly constructively well-defined; 3) a detailed map of the continuum phase structure of the same model(s); 4) a detailed understanding of the effective cosmological equations for the very early Universe emerging from the fundamental quantum gravity dynamics so encoded, and a quantum gravity solution to cosmological puzzles (flatness and horizon problems, the cosmological constant, the cosmological singularity) within the full theory, including a theory of cosmological perturbations from first (quantum gravity) principles, to be tested agains observations. This is an ambitious programme, as any quantum gravity programme has to be. The main asset of the GFT formalism, we believe, is the novel look it provides on canonical LQG as well as on spin foams, and the new tools we mentioned, with the simultaneous possibility to incorporate and take advantage of all the nice and important results obtained in such contexts, alongside those obtained in related formalisms like tensor models. It is this fruitful blend of solid, established results and techniques coming from different corners with a novel, promising new perspective that we hope will guarantee many more results and scientific surprises in the future, and a decisive progress for the whole field of quantum gravity.

\end{document}